# Lithium in Very Metal-poor Dwarf Stars - Problems for Standard Big Bang Nucleosynthesis?

David L. Lambert

*The W.J. McDonald Observatory, University of Texas, Austin, Texas, USA*

**Abstract.** The standard model of primordial nucleosynthesis by the Big Bang as selected by the WMAP-based estimate of the baryon density ($\Omega_b h^2$) predicts an abundance of $^7$Li that is a factor of three greater than the generally reported abundance for stars on the Spite plateau, and an abundance of $^6$Li that is about a thousand times less than is found for some stars on the plateau. This review discusses and examines these two discrepancies. They can likely be resolved without major surgery on the standard model of the Big Bang. In particular, stars on the Spite plateau may have depleted their surface lithium abundance over their long lifetime from the WMAP-based predicted abundances down to presently observed abundances, and synthesis of $^6$Li (and $^7$Li) via $\alpha + \alpha$ fusion reactions may have occurred in the early Galaxy. Yet, there remain fascinating ways in which to remove the two discrepancies involving aspects of a new cosmology, particularly through the introduction of exotic particles.

## INTRODUCTION

Observers and theoreticians have long regarded Big Bang nucleosynthesis as providing a simple testable set of reliable predictions with one free parameter to be determined by observations of the primordial relative abundances of the light nuclides $^1$H, $^2$H, $^3$He, $^4$He, and $^7$Li. That free parameter is the baryonic density $\Omega_b h^2$ or, equivalently, the number density ratio of baryons to photons $\eta$: $\Omega_b h^2 = 3.652 \times 10^7 \eta$ where $h$ is the Hubble constant in units of 100 km s$^{-1}$ Mpc$^{-1}$. Once the lifetime of the neutron and the number of neutrino families were known accurately from terrestrial experiments, the standard model of the Big Bang with the known nuclear reaction rates for the limited set of reactions controlling primordial abundances gave rather precise predictions for the abundance ratios of the light nuclides. An observer's goal was to establish the primordial abundance ratios (i.e., X/H) and to determine if each ratio gave the same value for $\Omega_b h^2$.

Many observers and theoreticians have believed for years that the various observed ratios did indeed indicate that one value of $\Omega_b h^2$ could satisfy the inferred primordial abundance ratios. This consistency involving abundance ratios differing by ten orders of magnitude was commonly cited as evidence for the standard Big Bang. Yet, some investigators emphasised apparent inconsistencies between the $\Omega_b h^2$ values provided by the different abundance ratios. Some ascribed the inconsistencies to systematic errors in the abundances derived from astronomical spectroscopy. Others chose to question aspects of the standard Big Bang and its attendant nucleosynthesis. These doubters of standard Big Bang nucleosynthesis were few in number.

With the mapping of the cosmic microwave background and the measurement of the acoustic peaks, the critical quantity $\Omega_b h^2$ has been determined accurately and quite independently of the primordial abundances. Publication of the accurate WMAP results (Spergel et al. 2003) has led to a resurgence of interest in primordial nucleosynthesis – see, for example, Olive's paper in this volume. In particular, attention has been given to the discrepancy between the lithium ($^7$Li) abundance predicted from standard Big Bang nucleosynthesis for the $\Omega_b h^2$ from the WMAP analysis and the lower value inferred for the primordial lithium abundance from observations of the Li I lines in very metal-poor dwarf stars. This discrepancy is a considerable factor of three in the number ratio $^7$Li/H. Measurements of $^6$Li in some very metal-poor dwarfs indicate an abundance of this isotope which is about a thousand times that predicted by the WMAP selected model of the Big Bang. These two discrepancies are the focus of this paper.

# STANDARD PREDICTIONS

Several recent studies of standard primordial nucleosynthesis have made new predictions of the abundances of the light nuclides. In light of the fact that the uncertainties in the standard suite of nuclear reaction rates dominate the uncertainties of the predicted abundances, these studies to differing degrees assess the accuracy of the reaction rates and apply mathematical tools to assess the uncertainties of the predicted abundances. For our purpose, it suffices to report the predictions without detailed comment on the underlying data on the nuclear reaction rates. Below, I give the authors' predictions for $\Omega_b h^2 = 0.0224 \pm 0.0009$ (Spergel et al. 2003), which is the WMAP result assuming a varying spectral index for the primordial fluctuations and considering other cosmic microwave background measurements on angular scales smaller than those well sampled by WMAP, and other kinds of observations (the Ly$\alpha$ forest, 2dF redshifts) for large angular scales. Abundances of D ($\equiv{}^2$H), $^3$He, and $^7$Li are given with respect to $^1$H as a ratio of number densities. By convention, the $^4$He abundance is given as a mass fraction $Y_p$.

Three sample predictions are:

- Coc et al. (2004) predict D/H = $(2.60^{+0.19}_{-0.17}) \times 10^{-5}$, $Y_p = 0.2479 \pm 0.0004$, $^3$He/H = $(1.04 \pm 0.04) \times 10^{-5}$, and $^7$Li/H = $4.15^{+0.49}_{-0.45} \times 10^{-10}$.
- Cuoco et al. (2004) predict D/H = $(2.55^{+0.22}_{-0.39}) \times 10^{-5}$, $Y_p = 0.2483^{+0.0008}_{-0.0005}$, $^3$He/H = $(0.99^{+0.07}_{-0.08}) \times 10^{-5}$, and $^7$Li/H = $4.9^{+1.4}_{-1.2} \times 10^{-10}$. These authors adopt $\Omega_b h^2 = 0.023^{+0.003}_{-0.002}$ from their analysis of the WMAP power spectrum.
- Cyburt (2004) predicts D/H = $(2.55^{+0.24}_{-0.19}) \times 10^{-5}$, $Y_p = 0.2485 \pm 0.0005$, $^3$He/H = $(1.011^{+0.074}_{-0.073}) \times 10^{-5}$, and $^7$Li/H = $4.27^{+1.02}_{-0.83} \times 10^{-10}$.

The mix of ingredients which define the standard Big Bang cosmological model and its attendant nucleosynthesis would be validated were these predicted abundances to be confirmed by the primordial abundances inferred from observations. Here, it must suffice to note that

- The D/H measured from high redshift absorption line systems in QSO spectra seems entirely consistent with the predictions. Kirkman et al.'s (2003) best estimate from five QSOs is D/H = $(2.78^{+0.44}_{-0.38}) \times 10^{-5}$. Dissenters exist but, if this measurement is accepted as the primordial value, observation and prediction are in excellent agreement.
- Observational data on the $^3$He abundance are scant. What data are available on $^3$He/H or $^3$He/$^4$He must be corrected for contamination of the observed gas by ejecta from earlier generations of stars. The magnitude and even the sign of the correction are uncertain. From their observations of Galactic H II regions, Bania, Rood, & Balser (2002) recommend $^3$He/$^4$He = $(1.1 \pm 0.2) \times 10^{-5}$ as the upper limit to the primordial abundance, a value quite consistent with the prediction according to the WMAP value of $\Omega_b h^2$.
- The predicted $Y_p$ exceeds all published determinations based on the emission line spectroscopy of Galactic and extragalactic H II regions and extrapolation of the He abundances to zero oxygen abundance. As an example, I note that Izotov & Thuan (2004) give $Y_p = 0.2421 \pm 0.0021$ from spectra of 82 H II regions in 76 blue compact galaxies with oxygen abundances ranging from 1/4 to 1/30 of the solar value. This difference of 2−3% between observation and prediction is a fragile basis for drastic redesign of the cosmological model. An evaluation of systematic effects in the abundance analyses deserves close scrutiny ahead of construction of non-standard Big Bangs. Recently, Olive & Skillman (2004) attempt 'a realistic determination of the error' and, after evaluating published observational determinations, offer $Y_p = 0.249 \pm 0.009$ as a representative value, but argue that allowed values are in the range $0.232 \leq Y_p \leq 0.258$. These assessments are in line with the WMAP-based prediction.
- In sharp contrast to the situation with D, $^3$He, and $^4$He, the predicted and observationally-inferred primordial abundances of $^7$Li are in sharp disagreement. The lithium abundance is derived from Li I lines in spectra of very metal-poor warm dwarf stars. Lithium in these stars has long been known to be dominated by the $^7$Li isotope (Maurice, Spite, & Spite 1984). For a representative measurement of the primordial abundance, I take the result obtained by Ryan et al. (2000): Li/H = $(1.23^{+0.68}_{-0.32}) \times 10^{-10}$. This is a factor of three less than the prediction based on the WMAP estimate of the baryonic density. Predicted abundances of $^6$Li are exceedingly low: $^6$Li/$^7$Li $\sim 10^{-5}$, but $^6$Li has recently been detected in some very metal-poor stars at a level of about $10^{-2}$.

This comparison of predicted and observationally-inferred abundances points to interesting discrepancies for $^7$Li, the least abundant of the five standard nuclides, and for $^6$Li. These discrepancies are discussed here. Do they imply that the standard picture of primordial nucleosynthesis needs modification? Or is it that the true primordial abundances of $^6$Li and $^7$Li differ from those abundances identified by observers as the primordial values?

# THE LITHIUM DISCREPANCIES - RESOLUTIONS?

Resolution of the lithium discrepancies has focussed on one or more of the following propositions:

- The nuclear reaction network adopted for primordial nucleosynthesis is incomplete and/or errors in the adopted reaction rates result in a systematic overestimate of the predicted $^7$Li abundance, and/or gross underestimate of the $^6$Li abundance.
- Systematic errors affect the determinations of the $^6$Li and $^7$Li abundances of very metal-poor stars.
- The $^6$Li and $^7$Li abundances of very metal-poor stars are not those of the primordial gas. In particular, the lithium abundance in the atmosphere of a very metal-poor star may have been reduced below its value in the natal stellar clouds by processes within the star. Such processes (see below) alter the $^6$Li/$^7$Li ratio but are not expected to elevate the $^6$Li abundance by large factors. The clouds may have been seeded by $^6$Li (and $^7$Li) produced by collisions between high energy $\alpha$s and ambient $\alpha$s.
- The standard physics incorporated into the Big Bang nucleosynthesis predictions is incomplete.

In the following sections, I comment on these four propositions.

# NUCLEAR REACTION RATES

Cross-sections for ten of the key reaction rates controlling synthesis of D, $^3$He, $^4$He, and $^7$Li have been measured at the energies relevant to the episode of Big Bang nucleosynthesis. Available data on the ten principal reactions are sufficiently accurate that the predicted primordial lithium abundance cannot be reduced by the required factor of about three to match the observed abundance attributed to metal-poor dwarf stars – see, for example, the above three cited references for the predicted abundance and its error bars. (The two reactions for which theory provides the necessary data are $n \leftrightarrow p$ and $p(n,\gamma)^2$H.)

The uncertainty of the predicted $^7$Li abundance at $\Omega_b h^2$ values around the WMAP value is dominated by the uncertainties over the rate for $^3$He($\alpha, \gamma$)$^7$Be. The $^7$Be through electron capture is the source of much of the primordial $^7$Li. The final $^7$Li abundance scales approximately linearly with the rate constant, and the quoted uncertainties are almost completely dominated by the adopted error in this rate constant. The fact that this reaction followed by $^7$Be($p, \gamma$)$^8$B affects the Sun's flux of $^8$B neutrinos led Cyburt, Fields, & Olive (2004) to use the Sudbury Neutrino Observatory's measurement of the neutrino flux in conjunction with a standard model of the solar interior to estimate the maximum allowable reduction of the $^3$He($\alpha, \gamma$)$^7$Be rate. Even this 'solar' rate, which is significantly smaller than the lower limits set by the various theoretical and experimental direct investigations by nuclear physicists, corresponds to a prediction of the lithium abundance from the Big Bang at the WMAP value of $\Omega_b h^2$ – $^7$Li/H = $(2.72^{+0.36}_{-0.34}) \times 10^{-10}$ – which eases but does not eliminate the lithium discrepancy.

Coc et al. (2004) report on a search for reactions not generally considered in the primordial synthesis of lithium. They suggest that $^7$Be($d, p$)$2^4$He might provide destruction of $^7$Be at a rate sufficient to remove the lithium discrepancy. The reaction rate listed in the standard compilation (Caughlan & Fowler 1988) is based on a single set of experimental data at energies greater than the Gamow peak for conditions of primordial nucleosynthesis. Coc et al. note that, if the extrapolation of the rate to lower energies is increased by a factor of about 300, the lithium discrepancy is removed. They further comment that the large increase is 'not supported by known data, but considering the cosmological or astrophysical consequences, this is definitely an issue to be investigated.' Coc et al. note that $^7$Be($d, \alpha$)$^5$Li is possibly another overlooked way to reduce the $^7$Li abundance. No data exists for this reaction.

In summary, the possibility that the lithium ($^7$Li) discrepancy will be resolved by revision of the reaction network and its rates is very slight. The one possibly open question should be resolved shortly by new measurements of the $^7$Be($d, p$)$2^4$He reaction.

Two reactions control the $^6$Li abundance from the Big Bang: production via $^2$H($\alpha, \gamma$)$^6$Li and destruction via $^6$Li($p, \alpha$)$^3$He (Thomas et al. 1993; Nollett, Lemoine, & Schramm 1997). Although the rate for the former reaction is relatively uncertain, it is clear that the predicted primordial $^6$Li/$^7$Li ratio must be so very small that detection of primordial $^6$Li from spectra of metal-poor stars is impossible. Cuoco et al. estimate $^6$Li/$^7$Li $\simeq 3 \times 10^{-5}$ for the WMAP-based $\Omega_b h^2$. Present understanding of the rates influencing the $^6$Li abundance suffices to conclude that primordial $^6$Li is undetectable.

# LITHIUM ELEMENTAL AND ISOTOPIC ABUNDANCES

That the cosmological potential of lithium may be realizable is due to Spite & Spite (1982) who showed that the lithium abundance for a small sample of warm metal-poor unevolved stars was independent of a star's metallicity. This lithium abundance as measured and identified by them as the primordial value was Li/H = $(1.12 \pm 0.38) \times 10^{-10}$, or, equivalently, $\log \varepsilon(\text{Li}) = 2.05 \pm 0.15$ on the usual astronomical scale where $\log \varepsilon(\text{H}) = 12.0$. Maurice, Spite, and Spite (1984) showed that the dominant isotope was $^7$Li: a limit $^6$Li/$^7$Li $< 0.1$ was set for a couple of stars.

In subsequent years, a voluminous literature has accumulated on the lithium abundance in metal-poor stars but the published abundances of lithium on the Spite plateau have varied little and none have neared the prediction corresponding to the WMAP-based $\Omega_b h^2$. Here, I comment briefly on some recent examinations of the Spite plateau, the dispersion in lithium abundances along the plateau, and the possibility that the lithium discrepancy may be erased by uncovering errors in the abundance analyses. In a later section, I discuss whether astrophysical effects (depletion, diffusion, etc.) might be held accountable for depressing the predicted lithium abundance to the lower observed value.

Lithium abundances are obtained from high-resolution spectra providing the Li I 6707Å resonance doublet in absorption. (The weaker line at 6104Å from the upper state of the resonance transition has been used to a limited exent - see below.) The basic atomic data for the 6707Å doublet are thoroughly well known from theory and experiment: wavelengths of the fine and hyperfine components for both $^6$Li and $^7$Li, and the $gf$-values. Synthetic spectra of the star are computed and matched to the observed Li I line. The syntheses use a model stellar atmosphere.

Of recent times, the baseline abundance analysis uses a classical model atmosphere computed according to the assumptions of plane-parallel homogeneous layers, hydrostatic equilibrium, local thermodynamic equilibrium (LTE), and flux constancy with the energy carried by a combination of radiation and convection. The Li I line is assumed to be formed in LTE. In analysing a star, one chooses the appropriate model atmosphere considering estimates of a star's effective temperature ($T_{\text{eff}}$), surface gravity ($g$), and composition. For unevolved stars which, of necessity, are used in the attempt to pin down the cosmological lithium abundance, the derived lithium abundance is sensitive to the choice of $T_{\text{eff}}$ and insensitive to the surface gravity and the assumed composition. Since the 6707Å line is weak for stars on the Spite plateau, the analysis is also only slightly sensitive to the adopted microturbulence and the damping constants. A bloodhound on the trail of systematic errors in the abundance analyses would sniff at the following leading suspects: the determination of $T_{\text{eff}}$ for metal-poor dwarf stars, the validity of LTE for Li I line formation, and the founding assumptions of classical atmospheres.

In the following subsections, I comment on three recent presentations of observations of the Li I doublet in metal-poor stars and the derivation of the inferred primordial lithium abundance. This is in no sense a critical and thorough review. The aim is simply to illustrate the consensus among most observers that there is a large gap between the inferred abundance and the prediction for a standard Big Bang corresponding to the WMAP estimate of $\Omega_b h^2$.

## Ryan and colleagues

Recent papers on primordial nucleosynthesis commonly cite the paper by Ryan et al. (2000), which is based on Ryan, Norris, and Beers (1999) for the source of the inferred primordial abundance. Stars contributing to this study span the metallicity range $-3.3 < $ [Fe/H] $ < -2.3$. The thrust of the 1999 paper is summarized by the paper's title 'The Spite Lithium Plateau: Ultrathin but postprimordial'. Here, 'ultrathin' means that the scatter in the lithium abundance is very small at a fixed metallicity, and 'postprimordial' refers to the slight increase in lithium abundance with increasing metallicity that is attributed to enrichment of Galactic gas with lithium. The primordial lithium abundance is identified with the extrapolation of the observed lithium abundances to zero metallicity. The 2000 paper estimates the primordial lithium abundance to be $\log \varepsilon(\text{Li}) = 2.09^{+0.19}_{-0.13}$, a value not significantly different from the original value given by Spite & Spite (1982). The estimate includes small ($<0.02$ dex) corrections for non-LTE effects affecting the Li I doublet and for depletion of surface lithium.

For the effective temperature scale, Ryan et al. (2000) adopt the Infra-red flux method (IRFM) calibration established by Alonso, Arribas, & Martínez-Roger (1996) which is about 120 K hotter than the scales considered by Ryan et al. (1999). This revision upward of the $T_{\text{eff}}$s increases the 1999 lithium abundances by 0.08 dex, and Ryan et al. (2000) consider that systematic errors of $\pm 120$ K may remain or the lithium abundances are uncertain to $\pm$ 0.08 dex from this source. In order that the inferred primordial lithium abundance be raised to the WMAP-based prediction the temperatures must be raised by about 900 K, an impossible systematic error. The need for such a severe increase could be reduced on the introduction of a correction to the IRFM $T_{\text{eff}}$s which increased with decreasing metallicity.

## Bonifacio and colleagues

In a recent study, Bonifacio et al. (2003, and private communication) adopt a $T_{\rm eff}$ scale based on a theoretical fit to the observed profile of the Balmer line H$\alpha$. The sample of 18 stars spans the metallicity range $-3.6 <$ [Fe/H] $-2.5$ with 11 stars having [Fe/H] $< -3.0$. When the published Li/H including corrections of 0.01 to 0.03 dex for non-LTE effects and standard depletion are plotted against Fe/H, a linear extrapolation by (my) eye to zero metallicity gives a primordial lithium abundance of $\log\varepsilon$(Li) = 2.20 with an uncertainty of about $\pm$ 0.1. This value is slightly higher than the Ryan et al. (2000) value, a difference very largely attributable to the higher $T_{\rm eff}$ scale from the H$\alpha$ profiles. (Bonifacio et al. quote a primordial abundance of 1.94 by extrapolation in the $\log\varepsilon$(Li) versus [Fe/H] plane.) The rate of the increase of lithium with metallicity is similar to Ryan et al.'s (1999). The scatter about a linear relation is small but slightly larger than that reported by Ryan et al. (1999).

## Asplund and colleagues

In an attack on the measurement of the isotopic $^6$Li/$^7$Li ratio in metal-poor stars on the Spite plateau (see below for comments on the isotopic ratios), Asplund et al. (2001, 2004 – in preparation) used the VLT/UVES combination to obtain very high-resolution ($R = \lambda/d\lambda = 10^5$), high S/N ratio (S/N $>$ 400 at 6707 Å) spectra for 24 stars with [Fe/H] between $-1.1$ and $-3.0$.

Lithium abundances obtained in a standard analysis with *MARCS* model atmospheres show again a positive trend of lithium with increasing metallicity. Extrapolation to zero oxygen (or iron) abundance gives $\log\varepsilon$(Li) = 2.09 when a quadratic fit to the full sample is made. The most oxygen-poor stars suggest a steeper decline of lithium with decreasing oxygen abundance: a linear fit to these stars gives $\log\varepsilon$(Li) = 2.03 on extrapolation to zero oxygen abundance. These values are from a $T_{\rm eff}$ scale based on a fit to the H$\alpha$ profiles and include the small corrections ($-0.02$ dex or so) for non-LTE effects but not the negligible correction for standard depletion. A $T_{\rm eff}$ scale based on photometry gives 50 - 100 K higher temperatures for the more metal-poor stars, or lithium abundances higher by 0.03 to 0.06 dex. The star-to-star scatter at a fixed metallicity is small for the H$\alpha$-based temperatures and even less than Ryan et al.'s (1999) scatter which lead to the label 'ultrathin'. A larger star-to-star scatter from the photometry-based temperatures using $V - K$ and $b - y$ is possibly due to errors in the adopted photometry and the uncertain corrections for interstellar reddening. The lithium abundances are less than those obtained by Bonifacio et al. (2003), also from a temperature scale based on H$\alpha$ profiles. This difference is presumably due to Asplund et al.'s use of an improved theory of line broadening for H$\alpha$. Asplund et al.'s lithium abundances are very similar to those given by Bonifacio & Molaro (1997) who used effective temperatures from the IRFM.

## The 6104 Å Li I line

A subordinate Li I line at 6104 Å has found limited use in the determination of the lithium abundance on the Spite plateau (Bonifacio & Molaro 1998; Ford et al. 2002; Asplund et al. 2004, in preparation). The lower level of this transition is the upper level of the resonance doublet. The 6104 Å line is weak so that high quality spectra are required for its accurate determination. Asplund et al. from their high quality UVES spectra (Figure 1) found no significant difference between the non-LTE abundances extracted from the 6707 Å and 6104 Å lines.

Ford et al. (2002) obtained identical non-LTE abundances of 2.19 from the 6707 Å and 6104 Å lines for HD 140283 for which they had a very high S/N spectrum; the LTE abundances differed by about 0.07 dex. The 6104 Å line is well-defined in their illustrated spectrum. For five other stars for which a detection of the 6104 Å was claimed, the non-LTE abundance from the 6104 Å line is about 0.3 dex higher than from the resonance doublet: the mean abundance for this quintet is 2.55, a value oddly close to the WMAP-based prediction. Examination of the published spectra of these stars shows a 6104 Å line which is rather ill-defined. Moreover, Ford et al. could set only upper limits on the lithium abundance for the other stars, all limits less than the quintet's mean. The authors write 'the large preponderance of upper limits prevent[s] any firm conclusion [about the difference in abundances from the 6707 Å and 6104 Å lines] being drawn.'

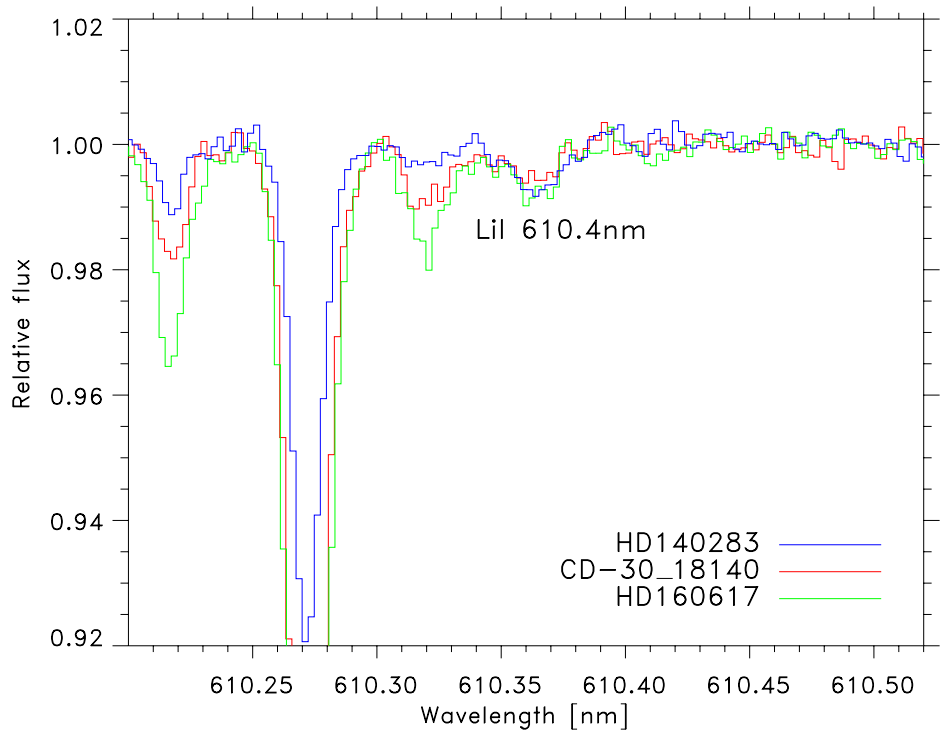

**FIGURE 1.** The region around the Li I 610.4 nm ($\equiv$ 6103.6 Å) for three metal-poor stars. The Fe I 610.22 nm, Ca I 610.27 nm, and Fe I 610.32 nm lines are present in this region. These lines have a different strength in each star. In contrast, the Li I line has a very similar strength in the three stars – an illustration of the Spite plateau.

## Beyond classical model atmospheres

The above three representative abundance analyses of the 6707 Å doublet used classical model atmospheres. Available grids of atmospheres – KURUCZ or MARCS models – give very similar results. Key assumptions, as noted above, behind construction of the grids are the representation of an atmosphere as a series of plane-parallel uniform layers in hydrostatic and local thermodynamic equilibrium with a constant flux carried through the atmosphere by the combination of radiation and convection. In short, if the stellar surface were resolved, it would be featureless with limb darkening the sole departure from uniform brightness.

The analyses, as noted, do include a correction for the departures from LTE in the formation of the Li I lines predicted by the LTE atmosphere. That inclusion of the correction is routinely made is due to the fact that (i) Carlsson et al. (1994) provided convenient tables of the corrections for the non-LTE, and (ii) the predicted corrections are small ($< |0.05|$ dex). The non-LTE calculations used a detailed model atom of 21 levels with radiative rates including 70 bound-bound and 20 bound-free radiative transitions, and with excitation and ionisation rates due to collisions with free electrons and hydrogen atoms considered. The published tables are exclusively for classical (1D) atmospheres.

Attempts are now being made to relax the classical assumptions in the construction of model atmospheres. In particular, the compressible radiative-hydrodynamics of the atmosphere are modelled. Then, the model atmosphere becomes a time-dependent three-dimensional (3D) construction. Models constructed for metal-poor dwarf stars are based on a code successfully applied to the Sun where solar granulation, asymmetric and Doppler-shifted absorption line profiles, and helioseismological observations are reproduced (Stein & Nordlund 1998).

An initial application of 3D models to the determination of the lithium abundance in metal-poor stars was made by Asplund et al. (1999) who reported that 'the primordial Li abundances may have been overestimated by 0.2 – 0.35 dex with 1D model atmospheres'. Calculations were limited to two representative stars: HD 84937 at the main sequence turn-off, and HD 140283 just beyond the turn-off. This analysis assumed the Li I lines were formed by LTE in the 3D atmospheres. (Model construction assumed LTE for both 1D and 3D cases.)

Asplund, Carlsson, & Botnen (2003) used the same 3D models but included non-LTE effects on the Li I line formation and found that the non-LTE effects restored the abundance close to the result from equivalent 1D atmospheres: the abundance from a 3D model was about 0.1 dex higher than the non-LTE abundance from the equivalent classical 1D model. This increase was reduced by Barklem, Belyaev, & Asplund (2003) who considered the effect on Li I level populations of the previously neglected charge-exchange process $Li(3s) + H \rightleftharpoons Li^+ + H^-$ with a small contribution from Li atoms in states other than 3s. The net effect is that the 3D non-LTE calculations with charge-exchange give an abundance for HD 84937 and HD 140283 within about 0.02 dex of that from the 1D LTE calculations. (The 1D non-LTE calculations by Carlsson et al. (1994) are changed by inclusion of charge-exchange collisions but corrections to the LTE abundances remain small: the correction which was +0.04 dex for HD 140283 becomes −0.05 dex, and the corresponding numbers for HD 84937 are −0.04 dex and −0.08 dex.)

Although one awaits continued refinement of the 3D models and associated line formation, these preliminary forays into application of 3D model atmospheres to the Spite plateau do not suggest that relaxation of the classical assumptions for model atmosphere construction will lead to a lithium abundance in accord with the WMAP-based prediction: a revision of the abundance based on 1D models upward by at least about 0.5 dex is sought but present 3D models with non-LTE line formation lower the derived abundance, albeit by only about 0.1 dex, and so aggravate the disagreement with the WMAP prediction.

## The lithium isotopic ratio

Standard models of the Big Bang predict an undetectable amount of the isotope $^6Li$, say $^6Li/^7Li \sim 10^{-5}$. Detection of $^6Li$ among stars inhabiting the Spite plateau then implies early synthesis of this isotope. Since $^6Li$ is destroyed about 70 times more readily by warm protons than $^7Li$, it has been assumed that the presence of $^6Li$ ensures that destruction of $^7Li$ by protons is negligible but this assumption is not necessarily valid (see below). The promise of learning about the Big Bang, the early history and even the prehistory of the Galaxy from measurements of the $^6Li$ abundance, and the window provided by $^6Li$ into processes of lithium depletion and diffusion has led to attempts to measure the isotopic ratio.

Standard models of metal-poor stars evolved from the pre-main sequence predict a slight depletion of $^6Li$ and negligible depletion of $^7Li$ for stars now at the main sequence turn-off. Predicted depletions increase for lower mass main sequence stars. Thus, the search for $^6Li$ has concentrated on stars around the main sequence turn-off. Presence of $^6Li$ is revealed by a wavelength shift to the red and an increased asymmetry of the Li I 6707 Å doublet (Figure 2). Detection of small amounts of $^6Li$ requires a good understanding of the factors – projected rotational velocity, microturbulence and macroturbulence – that shape the line profile. This understanding is based on fitting lines of other species (e.g., K I, Ca I, and Fe I) with a strength similar to that of the Li I doublet. High-resolution and high S/N spectra are a prequisite for a successful analysis. All reported analyses have been undertaken with classical 1D model atmospheres and the assumption of line formation by LTE. The observed asymmetry of the lines arising from stellar granulation is deemed too small to affect the determination of the lithium isotopic ratio – see, for example, Smith et al.'s (2001) discussion of the asymmetry of the K I 7699 Å resonance line in the spectrum of HD 84937, the star for which $^6Li$ was first detected. It should be noted nonetheless that 3D model atmospheres and their intrinsically asymmetric lines have yet to be applied to the isotopic abundance analysis.

The first detection of $^6Li$ in a metal-poor star was reported by Smith, Lambert, & Nissen (1993) for HD 84937 with $^6Li/^7Li = 0.06 \pm 0.03$. A non-detection of $^6Li$ was obtained for HD 19445, a lower mass main sequence star for which standard evolution predicts thorough destruction of $^6Li$. The search for $^6Li$ including reexamination of HD 84937 was continued by Hobbs & Thorburn (1994, 1997), and Smith, Lambert, & Nissen (1998). A higher-quality spectrum of HD 84937 was obtained and analysed by Cayrel et al. (1999) who found $^6Li/^7Li = 0.052 \pm 0.019$. A sample of five disk stars with [Fe/H] between −0.6 and −0.8 led to two positive detections of $^6Li$ with $^6Li/^7Li$ of 0.04 to 0.06 and three non-detections with $^6Li/^7Li < 0.01$ (Nissen et al. 1999). Asplund et al. (2001, 2004–in preparation) sought $^6Li$ in their high-quality spectra of 24 metal-poor stars (see above for discussion of their lithium abundances).

Results for the observed $^6Li$ and $^7Li$ abundances are presented in Figure 3. The stars with [Fe/H] < −1.7 from Figure 3 are shown in the Hertzsprung-Russell diagram in Figure 4 together with evolutionary tracks from VandenBerg et al. (2000) for two metallicities and three masses. It is clear that the sample is dominated by stars at and near the main sequence turn-off. There is a suggestion that the stars with detected $^6Li$ are closest to the turn-off, i.e., the hottest stars of the sample.

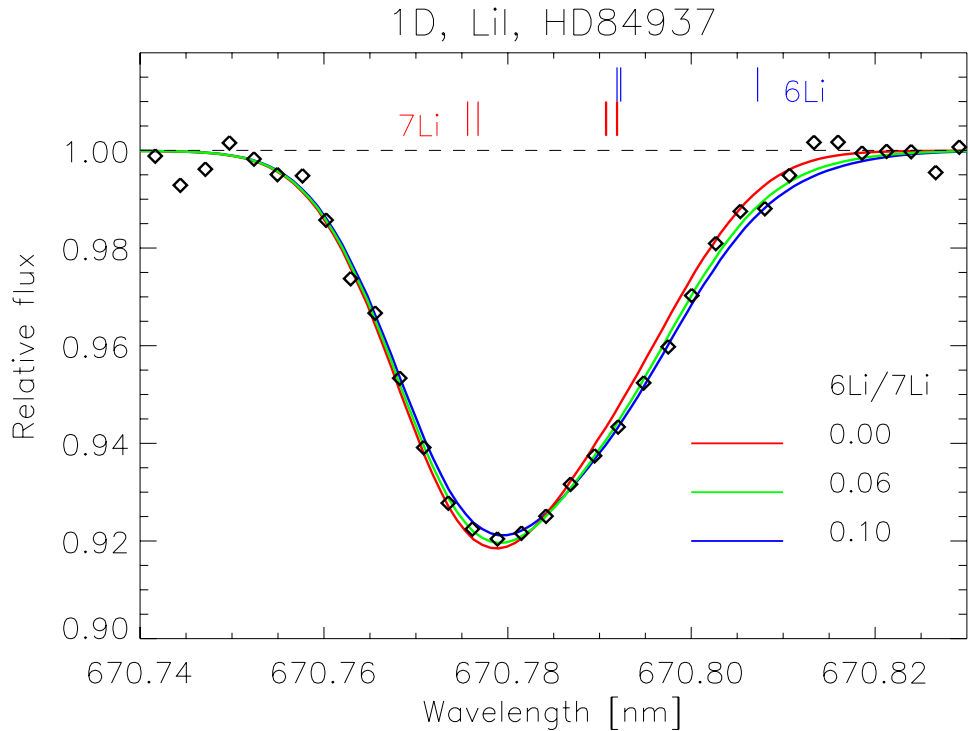

**FIGURE 2.** The Li I 6707 Å resonance doublet in HD 84937 from Smith et al. (1993). The wavelengths of the $^7$Li and $^6$Li components are indicated at the top of the figure. Synthetic profiles for three $^6$Li/$^7$Li ratios are shown – courtesy of Martin Asplund.

Figure 3 offers the hint that there is a $^6$Li plateau paralleling the Spite plateau but depressed below it by about -1.2 dex. All upper limits are at or below the $^6$Li plateau. The discussion later in the paper is predicated on the belief that the determinations of the $^6$Li abundances are without significant systematic error. Appearance of a $^6$Li plateau is possibly a challenge to the hubris of observers. The equivalent width and even the profile of the 6707 Å doublet vary little across the sample of observed stars, and, therefore, a systematic error in analysing the profile for a $^6$Li contribution could result in similar $^6$Li abundances for the sample and, hence, an apparent $^6$Li plateau. A partial counter to this scenario is the observation of no $^6$Li in HD 19445. Additional searches for $^6$Li among main sequence stars well below the turn-off would be valuable.

## Summary of observations

Spite & Spite (1982) described the lithium abundance of warm metal-poor dwarf stars by a single value: the abundance shared by all stars on the Spite plateau. The three selected recent analyses suggest that we may define the plateau more exactly with three quantities: the lithium abundance at zero metallicity, the slope in the $\log \varepsilon(\text{Li})$ versus metallicity (usually, [Fe/H]) plane at low metallicity, and the scatter in $\log \varepsilon(\text{Li})$ at a fixed low metallicity. To these three quantities concerning the elemental lithium abundance, I should add the presence of $^6$Li in some metal-poor stars.

The lithium abundance at zero metallicity is $\log \varepsilon(\text{Li})_0 = 2.1 \pm 0.1$, from the above discussions. All recent analyses show that the slope $d\log \varepsilon(\text{Li})/d[\text{Fe/H}]$ is positive: a value of around 0.1 dex is suggested. Scatter at a fixed metallicity is very small: $\sigma \simeq 0.03$ dex or less for the intrinsic scatter, after the observed scatter is corrected for the scatter from measurement errors. A few stars, which should inhabit the Spite plateau, show very large depletions of lithium, and are set aside in estimations of the scatter – none are shown in Figure 3. A very few stars show a lithium abundance clearly in excess of the plateau's value. One – HD 106038 – appears in Figure 3 with $\log \varepsilon(^7\text{Li}) = 2.48$ at [Fe/H] = $-1.35$. Another (not shown in Figure 3) is discussed by King, Deliyannis, & Boesgaard (1996): BD+23° 3912 with $\log \varepsilon(\mathrm{Li}) = 2.56$. A $^6\mathrm{Li}/^7\mathrm{Li}$ ratio of a few per cent is found among some turn-off stars.

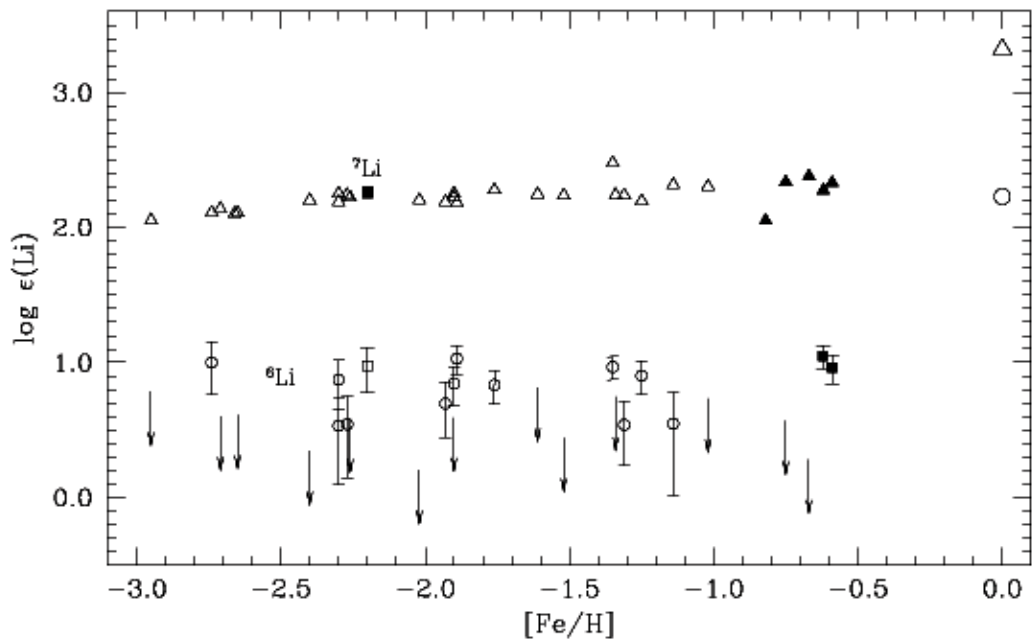

**FIGURE 3.** Observed abundances $\log \varepsilon(^7\mathrm{Li})$ and $\log \varepsilon(^6\mathrm{Li})$ as a function of [Fe/H]. Asplund et al.'s abundances are shown for $^7$Li (open triangles), and $^6$Li (open circles). Nissen et al.'s (1999) abundances are shown for $^7$Li (filled triangles) and $^6$Li (filled squares). Abundances for HD 84937 are shown by the filled square for $^7$Li and the open square for $^6$Li. Large symbols at [Fe/H] = 0 denote the solar system abundances: triangle = $^7$Li, and circle = $^6$Li.

One interpretation of these data identifies the lithium abundance at zero metallicity as the primordial lithium abundance, attributes the increase of lithium with [Fe/H] to Galactic synthesis of lithium including $^6$Li, and infers from the small star-to-star scatter in lithium abundance that the surface abundance cannot have been changed greatly from its initial value. One is left with a clearcut (0.5 dex) discrepancy between the inferred primordial lithium abundance and the WMAP-based prediction, i.e., a problem is posed for standard Big Bang nucleosynthesis.

An alternative interpretation accepts that the primordial lithium abundance was greater than $\log \varepsilon(\mathrm{Li})_0$ and stars on the Spite plateau have had their surface lithium abundance reduced over their lifetime of more than ten billion years. (The Sun at an age of 4.5 billion years and its 100-fold loss of surface lithium stands as a reminder that severe reductions occur, even if they remain ill-understood.) The challenge to advocates of this interpretation is to identify the mechanism(s) by which a reduction of 0.5 dex (assuming the primordial abundance was the WMAP value) can occur so uniformly in all stars. In this case, the slope $d\log \varepsilon(\mathrm{Li})/d[\mathrm{Fe/H}]$ may be a consequence of the combined effects of the depletion mechanisms and Galactic nucleosynthesis of lithium.

The commentary on the two contrasting interpretations of the Spite plateau serves as an introduction to the examination of depletion/diffusion mechanisms in the next section.

## DEPLETION AND DIFFUSION

A distinction is made between standard and non-standard theoretical predictions of the change of surface lithium abundance with stellar age. Standard here denotes stellar models without rotation, diffusion (gravitational settling, radiative levitation, .....), mass loss, and magnetic fields. Non-standard models include one or more of the physical effects neglected by standard models. It is useful here to recall Pinsonneault et al.'s (1999) admonition to sceptics that ' "Non-standard" does not imply "speculative"; in fact, there are solid physical grounds for including several effects neglected in the standard stellar model.' Vauclair's (2003) opening cry ' 'Microscopic diffusion' is a 'standard' stellar process" is an echo of Pinsonneault's admonition. In this section, I comment on a few recent papers that discuss non-standard models and their predictions for the Spite plateau. Each non-standard ingredient comes with a fairly extensive literature which the reader may trace from the provided references.

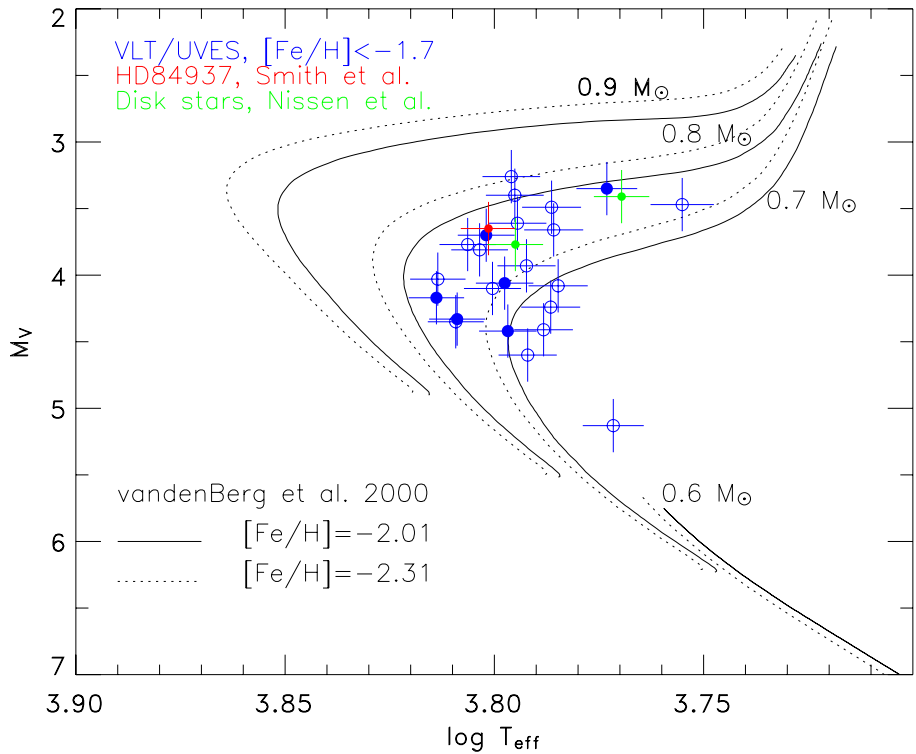

**FIGURE 4.** The Hertzsprung-Russell diagram for stars from Figure 3 with [Fe/H] < −1.7. Filled symbols denote stars with a detection of $^6$Li according to the key in the top left corner of the figure. Evolutionary tracks for the indicated stellar masses and metallicities are from VandenBerg et al. (2000).

## Standard models

In standard models, depletion of $^7$Li for metal-poor stars at and near the main sequence turn-off is negligible, i.e., less than 0.02 dex (Deliyannis, Demarque, & Kawaler 1990; Pinsonneault, Deliyannis, & Demarque 1992). Standard models explain quite well the decline in the lithium abundances for metal-poor stars with $T_{\text{eff}} < 5800$ K, the cool limit for the Spite plateau. In turn-off stars, surface $^6$Li is reduced by about 0.1 to 0.2 dex with most of this reduction occurring during the pre-main sequence phase. Predicted depletions of $^6$Li increase sharply for the main sequence stars below the turn-off.

If Nature's stars recognized only standard models and the effective temperature scale is without significant error, the observed lithium abundance (analyzed with 3D atmospheres and corrected for non-LTE effects) of stars on the Spite plateau could be taken as the primordial abundance, and the lack of dispersion on the plateau would be explained. The slope $d\log\varepsilon(\text{Li})/d[\text{Fe/H}]$ must be attributed to post-primordial nucleosynthesis of lithium which could include $^6$Li with $^7$Li.

## Atomic diffusion

In the case of the Sun, microscopic diffusion is a necessary ingredient in the solar model which reproduces the information on the interior provided by helioseismology: the run of the sound speed, the depth of the surface convection zone, and the He/H ratio at the surface. This success surely demands consideration of the effects of microscopic diffusion on the metal-poor stars defining the Spite plateau. Diffusion affects the interior and surface of a star. Sinking of helium in the interior reduces the stellar age. Diffusion of elements in the outer layers affects the surface abundances.

Atomic diffusion occurs in the presence of a concentration and temperature gradient. In the case of a star, the gravitational and radiation fields are key factors driving diffusion. The surface convection zone including the atmosphere is of a uniform composition. Diffusion occurs in the radiative zone below the thin convection zone. The base of the convection zone mixes with and attains the composition of the top of the radiative zone. Diffusion is a slow process but the metal-poor stars are old. Diffusion velocities decrease with increasing density. Thus, diffusion is least effective in the lowest mass stars where the convective envelope is largest and the density at its base the highest. Stars near the turn-off for which the convective envelopes are thin are most affected. Atomic diffusion leads to a reduction (and variation) of the lithium abundance throughout the radiative zone. Lithium is not supported by radiative levitation. Since the diffusion velocity decreases with increasing depth, the local lithium abundance increases with depth to reach a maximum at that depth where the timescale for downward diffusion equals the timescale for destruction of lithium by protons (Vauclair & Charbonnel 1998).

Early calculations (e.g., Michaud, Fontaine, & Beaudet 1984; Deliyannis & Demarque 1991) with atomic diffusion added to standard models revealed a problem in fitting the Spite plateau. The observed plateau is essentially flat for metal-poor stars hotter than about 5800 K, but models including atomic diffusion predict a significant drop in lithium abundance for stars hotter than about 6200 K, a drop not seen by observers. Setting aside the increased depletion predicted for the hottest stars, lithium in the stars on the truncated plateau at the presumed age of very metal-poor stars is predicted to be reduced by about 0.3 dex, an amount that partially closes the gap between observation and the WMAP-based prediction. Salaris & Weiss (2001) claim that available observations of the plateau are consistent with their predictions of lithium depletion driven by atomic diffusion provided that the metal-poor stars have an age between 13.5 and 14 Gyr. A key piece of this claim is the assertion that the number of observed stars is too few to define the predicted drop in lithium abundance for the turn-off stars. Recent observations (Bonifacio et al. 2003) of lithium contradict this assertion.

Richard et al. (2002, 2004 – also Michaud, Richard, & Richer 2004) introduce turbulent diffusion in the radiative zone to supplement the effects of atomic diffusion. A prescription is described for the turbulent diffusion coefficient. The successful prescription is far from an *ab initio* recipe. Addition of turbulent to atomic diffusion alters the predictions in two ways. First, the lithium depletion among the hottest stars is reduced so that the plateau can be extended across a wider temperature range than is possible with atomic diffusion acting alone. Second, mixing in the radiative zone destroys lithium and leads to lower surface lithium abundances across the plateau.

Richard et al. (2004) show that it is possible by a choice of the turbulent diffusion coefficient to reduce an assumed initial lithium abundance of 2.58 (i.e., essentially, the WMAP-based prediction) to 2.15, a value compatible with our recommendation for the observed abundance. This reduction is obtained over a temperature range covering the turn-off stars down to about 5900 K. It is not yet clear that the lack of scatter on the observed plateau is reproduced by the models; the simulation with turbulent diffusion was done for a single metallicity ([Fe/H] = $-2.3$) but a range of stellar masses and ages centered on 13.5 Gyr. Richard et al. (2002) report a simulation of the Spite plateau using atomic diffusion (no turbulence) over a range in metallicity, a dozen evolutionary tracks (i.e., masses), and a Gaussian distribution of ages centered on 13.5 Gyr with a 1.0 Gyr standard deviation. The dispersion about the plateau is very small and compatible with observations. Presumably, the turbulence may be related to rotation and a star-to-star difference in angular momentum history could then be expected to result in a dispersion of lithium abundances among turn-off stars.

In the case of atomic diffusion without turbulent diffusion, the $^6$Li/$^7$Li ratio for turn-off stars is increased slightly over its initial value, even after allowing for pre-main sequence depletion of lithium (Salaris & Weiss 2001; Richard et al. 2004). As implemented by Richard et al. (2004) turbulent diffusion of the strength necessary to bridge the gap between the WMAP-based prediction and observed lithium abundances leads to a larger destruction of $^6$Li than $^7$Li and, hence, a reduction of the surface $^6$Li/$^7$Li ratio below the initial value. The cited example that reduces the plateau abundance by 0.5 dex implies that an initial ratio $^6$Li/$^7$Li $\simeq 0.15$ is required for the lithium-depleted stars to have $^6$Li/$^7$Li = 0.05, as observed for HD 84937. This is a high value for an initial value - see later discussion on the synthesis of $^6$Li - and strongly incompatible with standard primordial nucleosynthesis.

Atomic diffusion affects the surface abundances of all elements. Richard et al. (2002) include predictions for 28 species from H to Ni. Predicted effects are largest, as expected, for turn-off stars and are reduced by the addition of turbulent diffusion. The abundance anomalies are weakened when a star evolves to become a subgiant and erased when the star becomes a giant thanks to growth of the convective envelope. Salaris & Weiss emphasized that the lithium abundance of subgiants rises to a maximum before declining for giants and that the maximum should be within about 0.1 dex of a star's initial lithium abundance, unless lithium is destroyed in the radiative zone, as may occur with turbulent diffusion. (Salaris & Weiss considered atomic diffusion without turbulent diffusion.)

Careful abundance analyses of turn-off stars may reveal unusual abundance ratios, say high ratios of (Mg, Al, Si, S) to (C,O). Since the abundance anomalies created by diffusion are erased by a giant's convective envelope, systematic differences in abundance ratios between dwarfs and giants would be a signature of diffusion-driven anomalies. There are no known striking differences for field stars. A valuable test of abundance anomalies induced by diffusion with and without a turbulent component would be provided by detailed abundance analyses of globular cluster stars from the main sequence up to the main sequence turn-off and beyond to the subgiants and giants. Salaris & Weiss and Richard et al. (2002) discuss the fragmentary data presently available.

## Rotationally-induced mixing

Stellar models that include mixing induced by rotation necessarily call for ingredients beyond those considered in the standard models (Pinsonneault et al. 1999): a prescription for the distribution of initial angular momenta; a recipe for loss of angular momentum; a recipe for the internal transport of angular momentum and associated mixing in radiative regions; the effect of rotation on the structure of the model. These prescriptions/recipes are not known from first principles but must be calibrated against the lithium abundances and rotational velocities of other stars, e.g., the Sun and the low mass stars in open clusters.

One recent attempt to model lithium depletion resulting from rotational mixing is by Pinsonneault et al. (1999) who construct models that reproduce the rotational velocities and lithium abundances of the Sun and stars in open clusters (Pleiades, Hyades, Praesepe, and M67). Lithium depletion of up to 3 dex in the low mass stars is primarily sensitive to assumptions about the Sun's initial angular momentum. Evolution of the Sun's angular momentum to its present internal distribution is used to calibrate the mixing coefficients for the radiative regions. The models predict an essentially flat Spite plateau for the stars at and just below the main sequence turn-off. (Rotationally-induced mixing would seem to be the sole non-standard process included in these models; atomic diffusion was apparently neglected.)

Dispersion in lithium abundance at a given evolutionary stage for a given mass and composition is a signature of depletion driven by rotational mixing. In Pinsonneault et al.'s calculations, the predicted dispersion ($\sigma$ in dex) scales with the $^7$Li depletion ($D_7$ in dex): $\sigma/D_7 \simeq 0.4$. Similarly, the depletions of $^6$Li and $^7$Li are correlated: $D_7/D_6 \simeq 0.4$. In standard models, $D_6$ is much greater than $D_7$, a reflection of the much larger destruction rate of $^6$Li by protons. Although destruction by protons is the dominant process for removing $^6$Li and $^7$Li in the models with rotationally-induced mixing, different fractions of material at the surface have been exposed to different temperatures and so different degrees of lithium destruction. In the limit that the surface is a mix of unexposed and severely exposed material, the $^6$Li/$^7$Li ratio is unchanged from its initial value, even as the lithium abundance declines.

The small dispersion reported by Ryan et al. (1999) and others (e.g., Asplund et al. 2004) implies a small depletion of lithium, as emphasised by these (and other) observers without an identification, much less an understanding, of the physical processes at work. As an illustration of the fact that their approach is not an *ab initio* one, Pinsonneault et al. use their result $\sigma/D_7 \simeq 0.4$ with the observed $\sigma$ to estimate $D_7$. Pinsonneault et al. (2002) conclude that 'Our best estimate of the overall depletion factor consistent with the RNB data is 0.13 dex, with a 95% range extending from 0.0 to 0.5 dex.' A depletion factor of 0.2±0.1 dex is recommended. Our estimate of $\log\varepsilon(\text{Li})_0 = 2.1$, if incremented by 0.2 remains less than the WMAP-based prediction, but the 95% range does encompass the prediction. A depletion of 0.2 dex for $^7$Li implies that the observed $^6$Li abundance be increased by 0.5 dex, and a star observed to have $^6$Li/$^7$Li = 0.05 would have initially had $^6$Li/$^7$Li $\simeq 0.15$, a value less than that ratio ($\sim 0.3 - 0.6$, see below) expected from production of $^6$Li (and $^7$Li) by $\alpha + \alpha$ collisions in a low density environment. The implied isotopic ratio would suggest a primordial $^7$Li abundance much less than the measured value for the Spite plateau. An initial $^6$Li/$^7$Li $\sim 1$ is suggested if the depletion is near the upper end of the 95% range. This ratio cannot be explained by collisions involving $\alpha + \alpha$ collisions. A radical change of primordial and/or pre-Galactic nucleosynthesis would be required in this latter case.

Vauclair (1999) noted that the settling of helium in the radiative zone leads to a vertical molecular weight ($\mu$) gradient. (The effects of the $\mu$-gradient were not considered by Pinsonneault et al.) Meridional circulation occurring in a rotating star sets up horizontal $\mu$-gradients. Mixing currents ensue that may cancel the effects of gravitational settling. Since lithium is fully-ionized in the radiative zone, it behaves like helium. If the effects of gravitational settling are cancelled by meridional circulation, Vauclair noted 'this could possibly account for the very small dispersion observed for the lithium abundances' on the Spite plateau. These ideas were explored by Théado & Vauclair (2001) in a paper titled 'On the possible existence of a self-regulating hydrodynamical process in slowly rotating stars'. Self-regulation of the lithium abundance for plateau stars was not demonstrated but 'under some conditions', it was shown that lithium depletion of a factor of two with a dispersion smaller than about 0.1 dex is expected. (This ratio of $D_7$ to $\sigma$ is similar to that provided by Pinsonneault et al. (1999).) Given the low limit now set of $\sigma$ by recent observations, it is not clear that this mechanism can be invoked to raise the observed lithium abundance by the required 0.5 dex to the WMAP-based prediction.

## Mass loss

Vauclair & Charbonnel (1995) considered the effect of a stellar wind on the surface lithium abundances of plateau residents. Mass loss was added to models in which atomic diffusion was included as a standard effect. As mass loss occurs, the depth of the surface convection zone remains essentially unchanged, but its base descends into the former top of the radiative zone in which lithium was diffusing inwards. If the base descends at lithium's diffusion velocity, the abundance of lithium in the envelope (and surface) is unchanged. On the other hand, a severe mass loss rate results in the convective envelope's base reaching deep into the radiative zone where lithium has been destroyed and the surface lithium abundance drops.

As noted above, atomic diffusion working alone produces a decrease in surface lithium abundance in the hottest stars: the predicted plateau has a turn down. To correct for this, mass loss rates must exceed $\dot{M} \simeq 10^{-13}\,M_\odot\,\mathrm{yr}^{-1}$ for these stars, and for all stars must be less than $2-5 \times 10^{-12}\,M_\odot\,\mathrm{yr}^{-1}$ in order that lithium not be decreased sharply. Differences in $\dot{M}$, age, and (presumably) metallicity result in changes to the surface lithium abundance, and, hence, a dispersion on the plateau. The dispersion appears to be small if stellar ages are less than 15 Gyr and $\dot{M}$ is in the range $10^{-12}$ to $-12.5\,M_\odot\,\mathrm{yr}^{-1}$. Lithium-6 is present at the surface provided that it was present in the star at its birth. (The few stars observed to have a $T_{\mathrm{eff}}$ of the plateau but very low lithium abundances may be stars with a much higher than average $\dot{M}$.)

Nothing is known about stellar winds from very metal-poor dwarf stars. The quoted mass-loss rates are too low for observational tests. The Sun experiences mass loss presently at about $10^{-14}\,M_\odot\,\mathrm{yr}^{-1}$. Deep short wavelength spectroscopy might detect or set useful limits on coronal lines from metal-poor dwarf stars.

## An observer's view

Although one may find a dissident or two among observers, the collective view is that there is a gap of about 0.5 dex between the lithium abundance of the Spite plateau and the WMAP-based prediction for standard primordial nucleosynthesis. The observed abundance is $\log\varepsilon(\mathrm{Li})_0 = 2.1\pm0.1$ with $^6\mathrm{Li}$ detected in several turn-off stars at the level $^6\mathrm{Li}/^7\mathrm{Li} \sim 0.05$. The WMAP-based predictions are $\log\varepsilon(\mathrm{Li})_0 = 2.65\pm0.1$ (simple average of the values cited in Section 2) and $^6\mathrm{Li}/^7\mathrm{Li} \sim 10^{-5}$. In this section, I have discussed ideas for bridging the gap by identifying mechanisms which may reduce the surface lithium abundance in plateau stars. None of the mechanisms, even if successful in bridging the gap, obviate the need for a source of $^6\mathrm{Li}$ other than a standard Big Bang.

An observer cannot but fail to be impressed by the variety and depth of study of processes affecting the surface lithium abundances of stars on the Spite plateau. Depletion of surface lithium abundance is predicted through operation of these processes, but achieving a depletion of 0.5 dex is a stretch for the majority of the processes. Yet, the challenge is not only to find a process – more likely, a combination of processes – that reduces the surface lithium abundance by the required 0.5 dex but does so with great uniformity for the observed stars on the plateau - with the exception of a few mavericks of very low lithium abundance. Perhaps, one may offer the 'atomic plus turbulent diffusion' models as today's 'best buy'. Then, the implied initial lithium abundance is probably consistent with the WMAP-based prediction when the various sources of error are taken into account. Given that atomic diffusion should be subsumed in the definition of a standard model, as now required for models of the solar interior, the only prescription of a non-standard ingredient is that for the turbulent diffusion. The maximum effects of this ingredient are partly constrained by the survival of $^6\mathrm{Li}$ in some stars and an assumption that the initial $^6\mathrm{Li}$ abundance was common among stars of the same metallicity and was not synthesized *in situ* by energetic particles (Deliyannis & Malaney 1995; Lambert 1995).

Determinations of the intrinsic dispersion of the lithium abundance for plateau stars now show it to be very small. It is difficult for many observers to accept that this is consistent with an overall reduction of 0.5 dex when that reduction is dependent on stellar properties, especially such variables as stellar rotation and mass loss rate. It is easy for observers to identify the lithium abundance of the plateau with the initial lithium abundance of early Galactic gas and to recognize that this abundance is not the WMAP-based prediction. In making this 'easy' identification, the rare examples of stars with a lithium abundance higher than the thickly populated plateau are dismissed. The known pair have abundances close to the WMAP-based prediction! Could these stars have retained more of their initial surface lithium abundance than the plateau population? Was the lithium enhancement created by mass transfer from an intermediate mass companion which synthesised lithium as a luminous AGB star?

A suspicion might be entertained that there are 'self-regulating' mechanisms awaiting discovery, as Vauclair (1999) has suggested. Yet, lithium is such a trace element that it is difficult to imagine that Vauclair's self-regulating mechanism which is likely heavily influenced by the most abundant elements (He, in particular) can lead to such uniform abundances for a trace fragile species. (Certainly, lithium burning cannot release significant amounts of energy.)

## SYNTHESIS OF $^6$LI

Synthesis of $^6$Li is attributed to spallation reactions (e.g., $p+$O $\rightarrow$ fragments, including $^6$Li, $^7$Li, Be and B) and to the fusion reactions $\alpha + \alpha \rightarrow ^6$Li and $^7$Li occurring when high energy particles collide with ambient particles in low density gas. Spallation was added to the inventory of nucleosynthesis mechanisms by Reeves, Fowler, & Hoyle (1970) with Galactic cosmic rays (GCR) providing the high-energy particles. The relative yields of $^6$Li, Be, and $^{10}$B were shown to be in good agreement with the solar system abundances. This was not the case for $^7$Li and $^{11}$B (relative to Be) whose abundances were higher than predicted. In the case of $^7$Li, stellar sources have been invoked to account for the increase of $^7$Li above the Spite plateau.

Spallation reactions are the sole identified process by which $^9$Be, the only stable isotope of beryllium, is synthesized. Synthesis of the boron isotopes $^{10}$B and $^{11}$B is also greatly impacted by spallation: $^{10}$B is exclusively a spallation product and $^{11}$B is likely so with the open possibility of a contribution from Type II supernovae via neutrino-induced spallation of $^{12}$C. Given the coupling between the synthesis of $^6$Li, Be and B, many discussions of the $^6$Li abundances in metal-poor stars also interpret the available Be and B abundances, which are now available for metal-poor stars and in some cases for the same stars for which $^6$Li has been sought.

Straightforward extension of GCR spallation to the metal-poor stars fails on two accounts. First, the ratio $^6$Li/Be = 5.6 for solar system material but very much higher ratios are found for plateau stars: e.g., $^6$Li/Be $\simeq$ 40 for HD 84937, and this ratio increases for some metal-poor stars with $^6$Li remaining approximately constant and Be declining with decreasing [Fe/H]. The initial ratios for these stars will have been higher because some $^6$Li depletion has likely occurred. Second, yields from GCR spallation are dominated by collisions between cosmic ray protons and $\alpha$s and interstellar C, N, and O nuclei. If this process had been dominant at all times, the Be abundances should scale approximately as Be/H $\propto$ (O/H)$^2$. Observations (Boesgaard et al. 1999) show, however, that Be/H $\propto$ (O/H) is a better description. Boron behaves similarly to Be (Duncan et al. 1997; García López 1998)). Duncan, Lambert, & Lemke (1992) suggested from an early description of the B $\propto$ O relation that the dominant channel for spallation involved collisions between fast C, N, and O nuclei and interstellar protons and $\alpha$s. If the fast C, N, and O are produced, as in supernova ejecta, with an abundance that is approximately independent of the initial metallicity of the exploding massive star, the yield of Be, and like products will be essentially independent of the metallicity, as is observed.

These ideas have been incorporated into the concept of superbubbles (Parizot & Drury 1999). Massive stars form in associations and clusters. The inevitable sequence of supernovae in rapid succession in a small region results in a superbubble of hot, tenuous gas with a composition dominated by the supernovae ejecta. The ejecta are likely mixed with ambient local material. The mixture is accelerated and undergoes spallation and fusion reactions. Models of light element nucleosynthesis in superbubbles and their contributions to galactic chemical evolution evolution of differ in their assumptions, particularly with respect to the composition and energy spectrum of the accelerated particles. It is considered that the mean energy of the particles will be less than that of the GCRs, a difference that affects the relative yields of light nuclides, especially of $^6$Li and $^7$Li relative to Be. A point of some controversy has been the energy budget – is the energy needed to create the spallation and fusion products comfortably less than the energy supplied as kinetic energy by the supernovae?

In Figure 5, I show predictions of the $^6$Li abundance as a function of [Fe/H] according to three different calculations. These calculations account satisfactorily for the observed Be and B abundances (not shown) and their run with [Fe/H]. The observed $^6$Li abundances are presumably lower limits to the initial $^6$Li values. Mercer et al. (2001) remeasured the cross sections for production of $^6$Li and $^7$Li in $\alpha + \alpha$ reactions finding $^6$Li production was ovestimated by previously adopted values by a factor of about two. This correction which is not included for the illustrated predictions may further foster the view that another source may be needed to account for the observed abundances of $^6$Li.

Production of $^7$Li via $\alpha + \alpha$ reactions may be invoked to account for the slope $d\log\varepsilon(\text{Li})/d[\text{Fe/H}]$ of the Spite plateau. If this identification is made, it is possible to predict the equivalent $^6$Li abundances. Adopting the lithium abundance near [Fe/H] $-3$ as the pre-Galactic abundance, the expected $^6$Li abundance at [Fe/H] $-2.5$ is $\log\varepsilon(^6\text{Li}) \simeq 0.7$ increasing to 1.3 at [Fe/H] $-1$, using Mercer et al.'s revised cross-sections for production via the fusion reaction. These expectations are close to the observed $^6$Li abundances which suggests that $^6$Li, where detected, is little depleted. The exercise may be reversed to argue that the spread in $^6$Li abundances and upper limits cannot represent the spread in initial abundances of this isotope because the spread implies a dispersion in $^7$Li abundances greater than is observed. If the Galactic production of $^7$Li comes from mass loss by luminous intermediate mass AGB stars which are known to make $^7$Li from a reservoir of $^3$He, there is no associated production of $^6$Li.

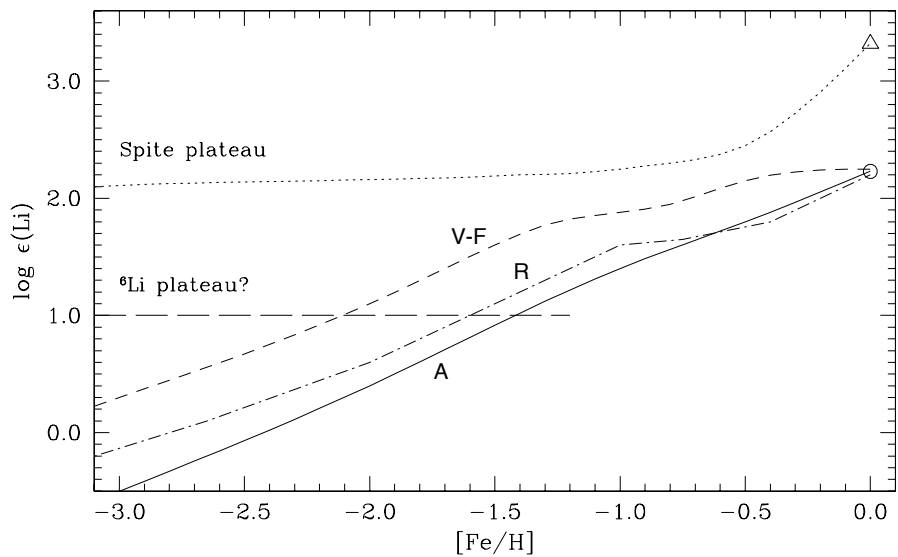

**FIGURE 5.** Predictions of the evolution of $^6$Li abundances with [Fe/H]. The Spite plateau for $^7$Li is represented by the sloping dashed line which is a fit by eye to the data in Figure 3. The line's extrapolation to the solar system abundance is not a fit to data. The evolution of the $^6$Li abundance is taken from Alibés et al. (2002, curve labelled A), Ramaty et al. (2000, curve labelled R), and Vangioni-Flam et al. (2000, curve labelled V-F). A possible $^6$Li plateau is suggested. Solar system abundances of the lithium isotopes are shown by the large symbols at [Fe/H] = 0.

The collection of $^6$Li measurements in Figure 3 appears to define a second plateau. Note that the measurements have not been corrected for depletion of $^6$Li in the pre-main sequence phase of evolution. Such corrections, when applied, necessarily push the implied $^6$Li abundances higher and may lessen the illusion of a plateau. Although prediction of absolute abundances of the light nuclides is a delicate and imprecise art, the general shape of the $^6$Li versus [Fe/H] relation expected from low energy energetic particles in superbubbles is not expected to differ greatly from those shown. Moreover, a similar shape and absolute abundances are found if the model of GCR protons and $\alpha$s spallating ambient C, N, and O is adopted (Ramaty et al. 2000). (Spallation by energetic particles in superbubbles and by GCRs in the interstellar medium, of course, result in very different predictions for Be (and B) versus [Fe/H].) Then, there would seem to be a possibility that a different origin of $^6$Li is needed at low [Fe/H] to create a plateau. A plateau implies that either the $^6$Li was a direct or an indirect product of the Big Bang or was produced early in the life of the Galaxy and unaccompanied by stellar nucleosynthesis.

Suzuki & Inoue (2002) propose that $\alpha + \alpha$ fusion reactions occurred as the result of gravitational shocks induced in the assembly of the Galaxy by infalling and merging subgalactic clumps. If the 'structure formation' shocks occur early and are limited in duration, the $^6$Li abundance is is built up and maintained at a constant level until production from superbubbles or GCRs pushes the abundance higher. Suzuki & Inoue propose a galaxy model in which shocks began at about 0.1 Gyr and lasted about 0.1 Gyr resulting in a rapid rise in $\log\varepsilon(^6\mathrm{Li})$ at about [Fe/H] = $-3$ to a plateau of $\log\varepsilon(^6\mathrm{Li}) = -1$ extending to [Fe/H] $\simeq -1.5$. This model crafted to fit the observation of $^6$Li in HD 84937 is deemed consistent with models of structure formation in the early Universe and energy requirements for lithium production by fusion reactions in shocks.

Various modifications of the standard prescription of primordial nucleosynthesis have been proposed for the two problems linked to lithium: an observed $^7$Li abundance on the Spite plateau at a level less than the WMAP-based prediction, and the presence of $^6$Li with an abundance possibly too high for attribution to sites other than the Big Bang. An observer of lithium in metal-poor stars who may regard the various modifications as drawn from a Pandora's box is likely nonetheless to be fascinated by putative links between stellar lithium abundances on the one hand and the Big bang and elementary particle physics on the other hand. One proposed link will be mentioned to conclude this section.

The standard model of particle theory is brilliantly successful but incomplete. Extensions of the model predict the existence of various exotic particles. If one or more of such exotic particles were to decay radiatively and/or hadronically during or after the episode of primordial nucleosynthesis, the decay products could process the primordial products and change the final abundances. A pioneering investigation of such effects of late-decaying exotic particles was reported by Dimopoulos et al. (1988). Very recent studies include those by Ichikawa, Kawasaki, & Moroi (2004), Jedamzik (2004), and Kawasaki, Kohri, & Moroi (2004). Jedamzik's conclusion is that 'a weak non-thermal hadronic source shows good potential to resolve two current discrepancies [i.e., the above $^6$Li and $^7$Li problems] in nuclear astrophysics'. Gravitino, neutralino, and $Q$-ball may now enter a stellar spectroscopist's vocabulary.

## CONCLUDING REMARKS

The two problems for standard Big Bang nucleosynthesis posed by stellar abundances of lithium are now well defined. Measurements of lithium in stars on the Spite plateau, even when obtained using state-of-the-art 3D model atmospheres and non-LTE line formation, are starkly 0.5 dex less than expected from the WMAP-based prediction for a standard Big Bang. In the absence of a severe systematic error afflicting the derived abundances, removal of the problem requires either that the stellar surface abundances have been reduced by about 0.5 dex over the lifetimes of the stars, or that primordial abundances are not those predicted for standard Big Bangs. (Discarding the WMAP estimate of $\Omega_b h^2$ introduces a different set of problems.) Multiple ideas have been proposed for reducing the surface lithium abundances but all face the challenge of accounting for a dispersion of a mere 0.03 dex after a 0.5 dex reduction of lithium abundance. Loss of lithium at the surface by atomic diffusion in the subsurface radiative layers may meet this challenge. The second problem is posed by the presence of $^6$Li in some metal-poor stars. Present measurements of $^6$Li abundances can possibly be accounted for by pre-Galactic or early Galactic synthesis of both Li isotopes by $\alpha + \alpha$ reactions in dilute gas. If the measured $^6$Li abundance has to be increased upwards to account for lithium depletion, the implied initial $^6$Li/$^7$Li ratio may create a problem for primordial nucleosynthesis. There is a tantalising hint of a $^6$Li plateau which may point to the work of exotic unstable particles such as gravitinos and neutralinos. Before invoking $-$inos, additional surveys for $^6$Li in very metal-poor stars and thorough scrutiny of systematic errors afflicting the measurement of the isotopic ratio are desirable.

With the introduction of 3D model atmospheres and non-LTE effects on Li I line formation leading to only minor revisions of the abundances, it seems safe to assert that the plateau's abundance cannot be as high as the WMAP-based prediction. Yet, steps should continue to be taken to improve the calibration of stellar effective temperatures and to refine the 3D models for metal-poor dwarfs. Non-LTE effects might be introduced in the models' construction for all key contributors to the line and continuous opacity as well as the electron donors. Present calculations assume that non-LTE effects are achieved instantaneously, i.e., steady-state statistical equilibrium (SSSE) is attained instantaneously throughout the atmosphere. Although it adds another dimension of complexity, an assessment should be made of the timescales required to achieve SSSE relative to the flow velocities for the uprising and downflowing plumes of gas.

An observational complement to theoretical work is vital too. In this respect, the asymmetry and velocity shift of different absorption line profiles may be observed and compared with predicted profiles. Asplund et al. (1999) note that the asymmetries of observed lines for HD 140283 (Allende Prieto et al. 1999) are well reproduced by the predictions. (Classical 1D atmospheres predict symmetric line profiles.) Observational scrutiny of line asymmetries is relevant to the interpretation of the intrinsically asymmetric 6707 Å resonance Li I doublet, especially to its application to the detection of $^6$Li which at the low detected abundances merely enhances the doublet's asymmetry.

In view of the constraint imposed on mechanisms of surface lithium loss by the true dispersion on the plateau, it behooves observers to extend the sample of observed stars and consider with great thoroughness the precision of their abundance determinations. A large sample of stars across the relevant dimensions of mass, age, and metallicity should be useful for testing proposed processes of lithium depletion. The claim of a postprimordial increase of lithium with metallicity rests not just on the lithium abundances but also on the metal abundances. Use of 3D instead of 1D models and consideration of non-LTE effects on the metal (iron, oxygen, ....) abundance are required in order to justify the claim thoroughly.

The appealing role played by atomic diffusion, even if modified by turbulent diffusion, calls for observers to include precise abundance determinations for other elements with the lithium measurements. As noted above, a critical test should be made on main sequence and subgiant stars in globular clusters. To probe the main sequence stars through high quality high resolution spectra will require a telescope with the light gathering power of the *Giant Magellan Telescope*.

# ACKNOWLEDGMENTS

I thank Martin Asplund for providing three figures, Piercarlo Bonifacio for sending lithium abundances in advance of publication, Georges Michaud for permission to quote from a preprint and for corrections to a draft of this review, Marc Pinsonneault for a helpful discussion, Ivan Ramirez for making two figures. I am happy to acknowledge the contributions of Martin Asplund, Poul Erik Nissen, Francesca Primas, and Verne Smith who are co-conspirators represented by Asplund et al. (2004). My research into stellar compositions is supported by the Robert A. Welch Foundation of Houston, Texas.

# REFERENCES

. Alibés, A., Labay, J., & Canal. R. 2002, ApJ, 571, 326
. Allende Prieto, C., Asplund, M., García López, R.J., Gustafsson, B., & Lambert, D.L. 1999, A&A, 343, 507
. Alonso, A., Arribas, S., & Martínez-Roger, C. 1996, A&A, 313, 873
. Asplund, M., Carlsson, M., & Botnen, A.V. 2003, A&A, 399, L33
. Asplund, M., Lambert, D.L., Nissen, P.E., Primas, F., & Smith, V.V. 2001, in *Cosmic Evolution*, ed. E. Vangioni-Flam, R. Ferlet, & M. Lemoine, World Scientific, 95
. Asplund, M., Nordlund, Å, Trampedach, R., & Stein, R.F. 1999, A&A, 346, L17
. Bania, T.M., Rood, R.T., & Balser, D.S. 2002, Nature, 415, 54
. Barklem, P.S., Belyaev, A.K., & Asplund, M. 2003, A&A, 409, L1
. Boesgaard, A.M., Deliyannis, C.P., King, J.R., Ryan, S.G., Vogt, S.S., & Beers, T.C. 1999, AJ, 117, 1549
. Bonifacio, P., & Molaro, P. 1997, MNRAS, 285, 847
. Bonifacio, P., & Molaro, P. 1998, ApJ, 500, L175
. Bonifacio, P. et al. 2003, IAU Joint Discussion 15, 39
. Carlsson, M., Rutten, R.J., Bruls, J.H.M.J., & Shchukina, N.G. 1994, A&A, 288, 860
. Caughlan, G.R., & Fowler, W.A. 1988. Atomic Data & Nuclear Data Tables, 40, 283
. Cayrel, R., Spite, M., Spite, F., Vangioni-Flam, E., Cassé, M., & Audouze, J. 1999, A&A, 343, 923
. Coc, A., Vangioni-Flam, E., Descouvemont, P., Adahchour, A., & Angulo, C. 2004, ApJ, 600, 544
. Cuoco, A., Iocco, I., Mangano, G., Pisanti, O., & Serpico, P.D. 2004, astro-ph/0307213, to appear in Int. J. Mod. Phys.
. Cyburt, R.H., 2004, Phys. Rev. D, 023505
. Cyburt, R.H., Fields, B.D., & Olive, K.A. 2004, Phys. Rev. D, 6913519
. Deliyannis, C.P., & Demarque, P. 1991, ApJ, 379, 216
. Deliyannis, C.P., Demarque,P., & Kawaler, S.D. 1990, ApJS, 73, 21
. Deliyannis, C.P., & Malaney, R.A. 1995, ApJ, 453, 810
. Dimopoulos, S., Esmailzadeh, R., Starkman, G., & Hall, L.J. 1988, ApJ, 330, 545
. Duncan, D.K., Lambert, D.L., & Lemke, M. 1992, ApJ, 401, 584
. Duncan, D.K., Primas, F., Rebull, L.M., Boesgaard, A.M., Deliyannis, C.P., Hobbs, L.M., King, J.R., & Ryan, S.G. 1997, ApJ, 488, 338
. Ford, A., Jeffries, R.D., Smalley, B., Ryan, S.G., Kawanomoto, S., James, J.D., & Barnes, J.R. 2002, A&A, 393, 617
. García López, R.J., Lambert, D.L., Edvardsson, B., Gustafsson, B., Kiselman, D., & Rebolo, R. 1998, ApJ, 500, 241
. Hobbs, L.M., & Thorburn. J.A. 1994, ApJ, 428, L25
. Hobbs, L.M., & Thorburn. J.A. 1997, ApJ, 491, 772
. Ichikawa, K., Kawasaki, M., & Takahashi, F. 2004, astro-ph/0402522
. Izotov, Y.I., & Thuan, T.X. 2004, ApJ, 602, 200
. Jedamzik, K. 2004, astro-ph/0402344

- Kawasaki, M., Kohri, K., & Moroi, T. 2004, astro-ph/0402490
- King, J.R., Deliyanis, C.P., & Boesgaard, A.M. 1996, AJ, 112, 2839
- Kirkman, D., Tytler, D., Suzuki, N., O'Meara, J.M., & Lubin, D. 2003, ApJS, 149, 1
- Lambert, D.L. 1995, A&A, 301, 478
- Mercer, D.J., Austin, S.M., Brown, J.A., Danczyk, S.A., Hirzebuch, S.E., Kelley, J.H., Suomijärvi, T., Roberts, D.A., & Walker. T.P. 2001, Phy. Rev. C, 63, 065805
- Maurice, E., Spite, F., & Spite, M. 1984, A&A, 132, 278
- Michaud, G., Fontaine, G., Beaudet, G. 1984, ApJ, 282, 206
- Michaud, G., Richard, O., & Richer, J. 2004, Mem. Ser. Astro. It. 75, 339
- Nissen, P.E., Lambert, D.L., Primas, F., & Smith, V.V. 1999, A&A, 348, 211
- Nollett, K.M., Lemoine, M., & Schramm, D.N. 1997, Phys. Rev. C, 56, 1144
- Olive, K.A., & Skillman, E.D. 2004, astro-ph/0405588
- Parizot, E., & Drury, L. 1999, A&A, 349, 673
- Pinsonneault, M.H., Deliyannis, C.P., & Demarque, P. 1992, ApJS, 78, 179
- Pinsonneault, M.H., Walker, T.P., Steigman, G., & Narayanan, V.K. 1999, ApJ, 527, 180
- Pinsonneault, M.H., Walker, T.P., Steigman, G., & Narayanan, V.K. 2002, ApJ, 574, 398
- Ramaty, R., Scully, S.T., Lingenfelter, R.E., & Kozlovsky, B. 2000, ApJ, 534, 747
- Reeves, H., Fowler, W.A., & Hoyle, F. 1970, Nature, 226, 727
- Richard, O., Michaud, G., & Richer, J. 2002, ApJ, 580, 1100
- Richard, O., Michaud, G., & Richer, J. 2004, ApJ, submitted
- Ryan, S.G., Beers, T.C., Olive, K.A., Fields, B.D., & Norris, J.E. 2000, ApJ, 530, L57
- Ryan, S.G., Norris, J.E., & Beers, T.C. 1999, ApJ, 523, 654
- Salaris, M., & Weiss, A. 2001, A&A, 376, 955
- Smith, V.V., Lambert, D.L., & Nissen, P.E. 1993, ApJ, 408, 262
- Smith, V.V., Lambert, D.L., & Nissen, P.E. 1998, ApJ, 506, 405
- Smith, V.V., Vargas-Ferro, O., Lambert, D.L., & Olgin, J.G. 2001, AJ, 121, 453
- Spergel, D.N., et al. 2003, ApJS, 148, 175
- Spite, F., & Spite, M. 1982, A&A, 115, 357
- Stein, R.F., & Nordlund, Å. 1998, ApJ, 499, 914
- Suzuki, T.K., & Inoue, S. 2002, ApJ, 573, 168
- Théado, S., & Vauclair, S. 2001, A&A, 375, 70
- Thomas, D., Schramm, D.N., Olive, K.A., & Fields, B.D. 1993, ApJ, 406, 569
- VandenBerg, D.A., Swenson, F.J., Rogers, F.J., Iglesias, C.A., & Alexander, D.R. 2000, ApJ, 532, 430
- Vangioni-Flam, E., Cassé, M., & Audouze, J. 2000, Phys. Rept., 333-334, 365
- Vauclair, S. 1999, A&A, 351, 973
- Vauclair, S. 2003, Ap&SS, 284, 205
- Vauclair, S., & Charbonnel, C. 1995, A&A, 295, 715
- Vauclair, S., & Charbonnel, C. 1998, ApJ, 502, 372